\documentstyle[a4,12pt,psfig]{article}
\newcommand{\al}{\alpha}
\newcommand{\bt}{\beta}

\newcommand{\be}{\begin{equation}}
\newcommand{\ee}{\end{equation}}
\newcommand{\ba}{\begin{eqnarray}}
\newcommand{\ea}{\end{eqnarray}}
\newcommand{\de}{\delta}
\newcommand{\dd}{\partial}
\newcommand{\ga}{\gamma}
\newcommand{\db}{\bar{\partial}}
\newcommand{\tet}{e_z^{~a}}
\newcommand{\tec}{e_{\bar{z}}^{~a}}
\newcommand{\teo}{e_0^{~a}}
\newcommand{\ta}{\theta}
\newcommand{\ach}{\mbox{arccosh}}
\newcommand{\bz}{\bar{z}}
\newcommand{\ra}{\rightarrow}
\newcommand{\la}{\lambda}
\newcommand{\nn}{\nonumber}
\newcommand{\p}{\psi^\al(z)}
\newcommand{\pbar}{\bar{\psi}_\beta(z_o)}
\newcommand{\vs}{\vspace{0.2cm}}
\newcommand{\vsa}{\vspace{0.4cm}}
\newcommand{\vsb}{\vspace{0.8cm}}
\begin{document}
\begin{titlepage}
\begin{flushright}
THU95/24\\
hep-th/9510060 \\
October 1995
\end{flushright}
\vsa
\begin{center}
{\large\bf Gravity in 2+1 dimensions\vs\\
           as a Riemann-Hilbert problem\vsa\vsb\\}
           M.Welling\footnote{E-mail: welling@fys.ruu.nl}
           \vsa\vsb\\
   {\it Instituut voor Theoretische Fysica\\
     Rijksuniversiteit Utrecht\\
     Princetonplein 5\\
     P.O.\ Box 80006\\
     3508 TA Utrecht\\
     The Netherlands}\vsb\vsa\\
\end{center}
\begin{abstract}
In this paper we consider 2+1-dimensional gravity coupled to
N point-particles.
We introduce a gauge in which the $z$- and $\bar{z}$-components of the
dreibein field become holomorphic and anti-holomorphic respectively.
As a result we can restrict ourselves to the complex plane.
Next we show that solving the dreibein-field: $e^a_z(z)$ is equivalent to
solving the Riemann-Hilbert problem for the group $SO(2,1)$. We give the
explicit solution for 2 particles in terms of hypergeometric functions.
In the N-particle case we give a representation in terms of conformal
field theory. The dreibeins are expressed as correlators of 2 free
fermion fields and twistoperators at the position of the particles.
\end{abstract}
\end{titlepage}
It is well known that in 2+1 dimensions, Einstein gravity is a topological
theory.
One way of understanding this is by counting the degrees of freedom of the
gravitational field.
One finds 6 independent metric components, 3 first class constraints which
generate 3 gauge
transformations (coordinate transformations). These have to be fixed by gauge
conditions,
introducing 3 more constraints. So the gravitational field has no local degrees
of freedom
(there are no gravitational waves or gravitons). It is however possible to
introduce some
degrees of freedom by the introduction of point particles.
In the case all these particles are static, the metric tensor $g_{\alpha\beta}$
was solved by Deser, Jackiw and 't Hooft \cite{beginarticle}.
Also a geometrical approach was invented by 't Hooft to treat moving point
particles \cite{tHooft}.
Here it was proved that a configuration of moving
point particles admits a Cauchy formulation within which no closed timelike
curves (CTC)
are generated. Spinning particles however do generate
CTC's and this is the reason why we will not consider them in this paper.
Quantum-mechanically one hopes that this problem will be resolved.

To attack problems in
2+1 dimensional gravity there are 2 mainstreams. There are researchers who use
Wittens
Chern-Simons formulation \cite{witten} and there are researchers who stick to
the ADM-formalism.
In the CS-formulation, point particles are coupled to non-abelian gauge
fields (that take values in the group-algebra $iso(2,1)$) by introducing
"Poincar\'e-coordinates" $q^a(x)$,
that transform as poincare vectors under
local gauge transformations \cite{CS}. We feel that the geometrical meaning
of the variables in the Chern-Simons approach is less clear than in the ADM
formulation. We will therefore stick to the ADM-formulation
of gravity. Furthermore we will only consider the case of an open universe.

Although the classical theory of gravity in 2+1 dimensions is relatively well
understood, the first-quantized models are not all equivalent
\cite{6ways} and it is not certain that a consistent quantum theory exists.
To our opinion however it is still interesting to find out what produces these
problems, and see if it is connected with the fundamentals of Einstein-gravity
(covariance) or if it is just an artifact of the 2+1-dimensional model.
It might be especially interesting to see if one can find a consistent
S-matrix for scattering of quantum-particles. The scattering of 2 quantized
point particles was first studied by 't Hooft and later by Deser and Jackiw
\cite{scattering1}.
In the CS-formulation scattering of particles was studied by Carlip and later
by Koehler et al.\cite{scattering2}.
To our knowledge a second-quantized model for i.e. a scalar field coupled
to gravity has not yet been investigated. In such a model one could study the
ultraviolet behaviour
\footnote{In 2+1 dimensions gravity coupled to a scalar field is still
non-renormalisable because the coupling constant is dimensionfull.}
 of Einstein-gravity because one expects to have much better
understanding of the non-perturbative theory than in 3+1 dimensions.

In what follows we will restrict ourselves to the classical theory of gravity
in 2+1 dimensions, coupled to a set of moving point particles. The total mass
of the particles must
be such that the universe does not close. This is necessary for our gauge
choice to be valid for all times t.
In this case we also avoid the problem of time, because we have a boudary term
that will act as a
Hamiltonian. So we use the asymptotic coordinate t as our time-variable.

In this paper we start to review some well known aspects of general relativity
in
section 1. It is stressed that for an open universe we need a boundary term in
the
action which serves as a Hamiltonian for the reduced theory. The reduced theory
must
be obtained by putting the solutions for the metric $g_{\mu\nu}$  back into the
action.
In 2+1 dimensional gravity the gravitational field can in principle be
completely
integrated out because it has no dynamical degrees of freedom. In the next
section
we introduce a slicing condition in such a way that the area of the Cauchy
surface is
locally maximal. On that slice we then choose
conformal coordinates. In this gauge the coordinates are not multivalued as in
the "flat
coordinate-system" (in the flat coordinate-system the metric is everywhere
$\eta_{\mu\nu}$
except at the particles positions).
Another pleasant feature of this gauge is that the tetrad field $e_{\mu}^{~a}$
defined by
the relation $g_{\mu\nu}=e_{\mu}^{~~a}e^b_{\nu}\eta_{ab}$, in complex
coordinates, splits
into a holomorphic and a anti-holomorphic sector. Actually all information is
contained
in the field $\tet$, which is a holormophic function and can be studied on the
complex
plane. A residual gauge invariance are the conformal transformations and we can
use them
to bring 2 particles to definit positions (leaving infinity invariant). Another
pleasant
property of the coordinates is the fact that in the process of reducing the
action, only
the kinetic term of the matter variables together with the boundary term
survives. As
mentioned, the boundary term takes the role of the Hamiltonian for the matter
fields.
Next we solve in our coordinate system the gravitational field for a
stationary, spinning
particle with mass M, sitting at rest in the origin. Although this metric is
pathological
near the particle, the asymptotic behaviour of this solution is very usefull as
asymptotic
condition for the general solution. This makes sense because we expect that any
configuration
of particles with total mass M and total angular momentum J can be described by
an effective
center of mass particle with mass M and spin S=J.
In section 3 we refrase the problem of finding solutions for the tetrads $\tet$
into an old
mathmatical problem: The Riemann-Hilbert problem.
The Riemann-Hilbert problem deals with a vectorfield $\tet$, in general living
on a
Riemann-surface with N singularities, that has {\em prescribed} monodromy
properties.
In our case we will work on the Riemann-sphere and the monodromy must take
values in
SO(2,1). Actually it proves more convenient use the spinor representation of
SU(1,1)
and find a spinor $\zeta^\al(z)$ with the corresponding monodromy properties.
The Riemann-Hilbert problem can be attacked in 2 different ways. The first
method
one can use seeks a d'th order linear differential equation with N regular
singularities.
(d is the dimension of the representation used; 3 for SO(2,1) and 2 for
SU(1,1)).
Throughout the paper we stress the importance of the local behaviour of the
solution
near the singular points (particles) and at infinity. As an example of this
method we
solve the 2 particle case in terms of hypergeometric functions. The N-particle
case is
more conveniently studied using a first order linear matrix differential
equation where
the dimension of the matrices is d. We use a representation due to Miwa, Sato
and
Jimbo \cite{Sato} to represent the solutions to this equation in terms of
correlators
of 2 free fermions with twistoperator insertions on the positions of the
particles.
We check that the local behaviour  of this solution near the singular point
indeed
agrees with what one expects. As we move the singular point (for instance by
evolution)
the monodromy does not change. This implies that we can write down a set of
Schlesinger
deformation equations. They describe basically how the matrix differential
equation and
its fundamental matrix of (rationally independent) solutions change if we
change the
deformation parameters $a_i$ (i.e. the location of the particles). Next we
check that
we can reproduce the static solution of N particles at rest. It should be noted
that
this special example bears much resemblance with the Coulomb gas. Finally we
derive
some kind of Knizhnik-Zamalodchikov equation for the correlator of the
twistfields
(without the fermions).
\newpage
\section{The ADM-formalism}
\label{ADM}
We will work on space-times with the topology $\Sigma\times R$ where
$\Sigma$ is an open orientable 2 dimensional surface. In the ADM-formalism one
makes
a 2+1-split of the metric in the following way:
\footnote{Most of the expressions in this section can be found in
\cite{gravitation}}
\ba
ds^2 &=& g_{\mu\nu}dx^\mu dx^\nu \nn\\
     &=& (-N^2+g_{ij}N^iN^j)dt^2+2N_idx^i dt+g_{ij}dx^i dx^j
\label{3+1split}
\ea
In this paper roman indices in the middel of the alphabet (i,j,k,...) will
range over {1,2},
roman indices in the beginning of the alphabet (a,b,c,...) are flat indices and
run over {0,1,2} and
greek indices are space-time indices and run over {0,1,2}. The indices of 2
dimensional tensors are
raised and lowered by $g_{ij}$. This object is also the dynamical variable of
the theory. The lapse
function $N$ and the shift functions $N_i$ are Lagrange-multiplier fields and
as such not dynamical.
They determine the way the 2 dimensional surface is inbedded in the 3
dimensional space-time. The field
canonically conjungate to $g_{ij}$ is:
\be
\pi^{ij}=\sqrt{\gamma}(K^{ij}-K^l_lg^{ij})
\label{pi}
\ee
Here $\sqrt{\ga}$ is the square root of the determinant of the 2-dimensional
metric, $g^{ij}$ is the inverse of $g_{ij}$ ($g_{ij}g^{jl}=\delta^l_i$),
and $K_{ij}$ is the extrinsic curvature that measures how the normal to the
surface $\Sigma$
changes as we walk along that surface.
It is determined by:
\be
K_{ij}=\frac{1}{2N}(\dd_t g_{ij}-2D_{(i}N_{j)})
\label{K}
\ee
Here $D_i$ is the covariant derivative with respect to the 2-dimensional
surface defined by $g_{ij}$. The brackets around indices means that we
symmetrize the expression in those indices.\\
The Lagrangian for gravity coupled to matter in terms of these variables is:
\ba
L&=& L_g+L_M \label{lagtot}\\
L_g&=& \int~d^3x~\pi^{ij}\dd_t g_{ij}-NH-N^iH_i \label{laggrav}\\
H&=&\frac{1}{\sqrt{\ga}}(\pi^{ij}\pi_{ij}-(\pi^l_l)^2)-\sqrt{\ga}~^{(2)}R
\label{ham}\\
H_i&=&-2D_j\pi^j_i \label{hami}\\
L_M&=&-16\pi G\int d\tau\sum_{n}m_n~\sqrt{-g_{\mu\nu}x'^\mu_n
x'^\nu_n}\label{lagmat}
\ea
where $^{(2)}R$ means the intrinsic, 2-dimensional curvature scalar and
$x'=\frac{\dd x}{\dd\tau}$. The equations of motion derived from this
action are then the Einstein equations:
\be
G_{\mu\nu}=8\pi G T_{\mu\nu}
\label{einsteineq}
\ee
where
\be
T^{\mu\nu}=\int d\tau~\sum_n~m_n~\frac{x'^\mu_n x'^\nu_n}
{\sqrt{-g_{\mu\nu}x'^\mu_n x'^\nu_n}}\frac{\de^3(x^\mu-x_n^\mu(\tau))}
{\sqrt{-g}}
\label{energytensor}
\ee
The expression for the matter Lagrangian can also be written in a
first order form:
\be
L_M=-16\pi G\int d\tau~[p_a e^{~a}_\al x'^\al-\frac{\lambda}{2}
(\eta^{ab}p_a p_b+m^2)]
\label{lagmat2}
\ee
This form is more convenient for us as we will work with the "dreibein"
$e_\mu^{~a}$. Of course the metric is easily recovered from the dreibein
by the formula:
\be
g_{\mu\nu}=e_{\mu}^{~a}e_{\nu}^{~b}\eta_{ab}
\label{g-e}
\ee
$p_a$ Has the interpretation as the momentum of the particle in a flat
coordinate system. One immediately recognizes what a formidable task it is to
solve these equations in the general case of N moving particles in arbitrary
directions with arbitrary velocities. We will therefore attack the problem from
a different perspective.\\
Finally we want to remark that in the case of an open universe Regge and
Teitelboim
\cite{ReggeTeitelboim} found that one should add to the Lagrangian
(\ref{lagtot}) a surface term
that is needed to have a well defined variational derivative. This surface term
is then identified with the total energy contained in the universe.
In 2+1- dimensions this surface term is \cite{Ash}:
\be
E=\frac{1}{16\pi G}\int dx^2 ~\sqrt{\ga}^{(2)}R
\label{surfaceterm}
\ee

It is now interesting to perform a reduction of the system to the physical
degrees of freedom.
One has to insert the solution of the gravitational fields into the total
Lagrangian (\ref{lagtot}).
In the coordinates chosen by us, this implies that $L_g$ will vanish completely
and
the surface term becomes the Hamiltonian for the reduced system. So it is an
"effective" Hamiltonian that describes how the point particles interact.
The fact that there are no terms that survive in $L_g$ is clear if one
remembers that the gravitational field itself contains no degrees of freedom.
We will return to this point in the discussion.

\section{Gauge fixing and asymptotic conditions}

\begin{figure}[t]
\centerline{\psfig{figure=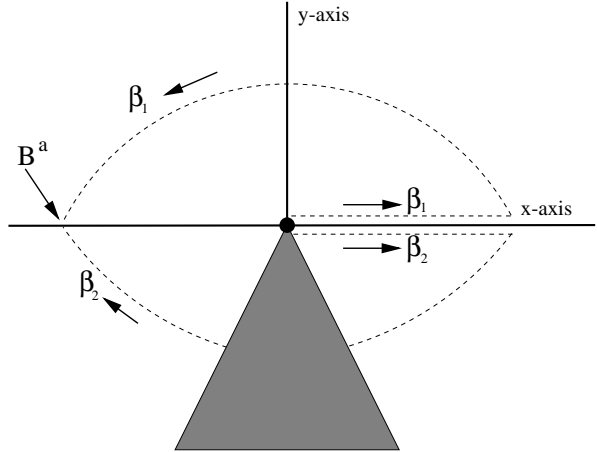,angle=-90,height=6cm}}
\caption{Space-time around a particle sitting at rest in the origin}
\label{wedge}
\end{figure}

The problem of deriving the gravitational fields around moving point particles
is already solved in the flat coordinate system \cite{tHooft}
in a geometrical approach. The flat coordinates $u^a$ are however multivalued.
One can see this already in the simplest configuration: a particle sitting
at rest in the origin. The space-time is constructed by cutting a wedge out of
space-time. Consider now a point $B^a$ somewhere in the plane (see
fig.\ref{wedge}).
If an observer approaches the point $B^a$ along path $\beta_1$ he will find
coordinates:
\be
B^a=\left( \begin{array}{ll}
            R\cos\varphi\\
            R\sin\varphi
              \end{array} \right)
\ee
However, an observer moving along path $\beta_2$ would give point $B^a$ the
coordinates:
\be
B^a=\left( \begin{array}{ll}
            R\cos(\varphi+\al)\\
            R\sin(\varphi+\al)
              \end{array} \right)
\ee
One observes that it is really the unusual ranges (or matching conditions) of
the coordinates
that make them multivalued. The main aim of this paper is to introduce
new, single-valued, coordinates $(z,\bz)$, that transform the flat metric
$\eta_{ab}$ into a nontrivial metric $g_{\al\bt}$ according to the well known
transformation law:
\be
g_{\mu\nu}=\dd_\mu u^a\dd_\nu u^b\eta_{ab}
\label{g-u}
\ee
Part of the gauge fixing-procedure is choosing a Cauchy-surface.
The Cauchy surfaces we are looking for, are conveniently
characterized by the property that they locally maximize the area of the
surface.
Since we know that
locally there must always be a surface that has maximal area, we know that a
solution to this
coordinate condition always exists (i.e. the gauge is accessible).
This is believed to be the simplest gauge condition that provides for single
valued
coordinates $(z,\bar{z},t)$, even though the elaboration of this condition is
fairly
complicated as will become clear.
The coordinates $u^a(z,\bz;t)$ are considered as embedding conditions for each
time t.
They are flat coordinates but unfortunately multivalued functions on these
Cauchy surfaces.
So in order to descibe them we need cuts in the surface, called strings in
previous literature.
The transformation of $u^a$ over a string is in general a Poincar\'e
transformation.
We can also define the so called dreibein or tetrad fields on the surface
$\Sigma$:
\be
e_\mu^{~a}=\dd_\mu u^a
\label{e-u}
\ee
They are also multivalued but transform by homogeneous Lorenztransformations if
we move over a string. In what follows we will often work in complex
coordinates:
\ba
z&=&(x+iy) \nn\\
\bar{z}&=&(x-iy)
\label{complex}
\ea
It is easy to write down an action that would produce the slicing condition
that
the area of the Cauchy surface $\Sigma$ is locally maximal. We can view the
coordinates $(z,\bz)$ as coordinates of a Euclidean worldsheet living in a flat
3 dimensional target-space described by the coordinates $u^a$. The embedding
conditions for the worldsheet to have maximal area are then precisely given by
the Polyakov action:
\footnote{In string theory one could say that the string follows a trajectory
so that the worldsheet has maximal area. Because we consider a Euclidean
"worldsheet" we can not speak about trajectories.}
\be
S_p=\int d^2z~\sqrt{\gamma}g^{ij}\dd_iu^a\dd_ju^b\eta_{ab}
\label{polyakov}
\ee
where $g_{ij}$ is the 2 dimensional metric on the worldsheet,
$\gamma=$det$g_{ij}$
and $\eta_{ab}$ is the 3 dimensional metric in the targetspace. We can also
view
the $g_{ij}$ as the induced metric on the Cauchy surface derived by projecting
$\eta_{ab}$ on $\Sigma$. If we furthermore choose these coordinates on the
worldsheet to be
conformally flat:
\be
g_{ij}=e^\phi\de_{ij}
\label{conformalmetric}
\ee
then this action reduces to:
\be
S_p=\int d^2z~\dd u^a\db u^b\eta_{ab}
\label{string}
\ee
where we used the complex coordinates $(z,\bz))$ and real coordinates $u^a$.
Furthermore we used the notation: $\frac{\dd}{\dd z}=\dd~~\frac{\dd}{\dd
\bar{z}}=\db$.
It is well known that the equations that follow from the string action are:
\ba
\dd u^a\dd u^b\eta_{ab}&=&0 \label{gauge1}\\
\db u^a\db u^b\eta_{ab}&=&0 \label{gauge2}\\
\dd\db u^a &=&0 \label{gauge3}
\ea
This implies that $u^a$ is a harmonic function:
$u^a(z,\bz;t)=u^a(z;t)+u^a(\bz;t)$.
These conditions imply for the dreibein field (\ref{e-u}):
\ba
e^{~a}_0\equiv \dd_t u^a &=& C^a(t)+\dd_t\int dz~\tet+\dd_t\int
d\bz~\tec\label{e0}\\
\tet\equiv \dd u^a &=&\tet(z)~~~\mbox{(holomorphic)}\label{ez}\\
\tec\equiv \db u^a &=&\tec(\bar{z})~~~\mbox{(anti-holomorphic)}\label{ebarz}\\
\tet e_z^{~b}\eta_{ab}&=&0~~~~~~~~~~~\mbox{(null-vector)}\label{nullez}\\
\tec
e_{\bar{z}}^{~b}\eta_{ab}&=&0~~~~~~~~~~~\mbox{(null-vector)}\label{nullebarz}
\ea
where $C^a(t)$ is some vector only depending on t.
Note that we still have a remaining conformal freedom: $z\rightarrow w(z)$.
This will be used later to transform 2 particles to definit positions (leaving
infinity invariant).
Using equations (\ref{conformalmetric}),(\ref{g-e}),(\ref{K}) and (\ref{pi})
one can calculate what
the implications are for the fields $g_{ij},\pi^{ij}$.
\ba
g_{ij} &=& e^\phi \de_{ij}\label{conformalmetric2}\\
\pi\equiv\pi^l_l&=&0\label{piis0}
\ea
The advantage of this gauge choice is now evident. First of all, as we will see
later, we will only have to solve the holomorphic, null-vectorfields $\tet$,
living on the
complex plane, in order to find the gravitational field $g_{\al\bt}$. The fact
that the
dreibein splits into a holomorphic and a anti-holomorphic part is a peculiar
fact of
2+1 dimensional gravity. From
a mathematical viewpoint it is a convenient gauge condition because we have now
the
machinery of complex calculus on Riemann surfaces at our disposal. The
conformal
freedom also suggests that there might be a connection with conformal field
theory.
As we will show later, the dreibein is formally solved as a two-point function
of
free conformal fermion fields with some twistoperator
insertions. Another advantage is that in the reduction scheme explained in
section \ref{ADM}, the reduced "effective" Lagrangian looks very simple.
Since $\pi$ is the canonicacally conjugate momentum of $2e^\phi$ and we have
formulas
(\ref{conformalmetric2}) and (\ref{piis0}) we find that in the Lagrangian
(\ref{laggrav})
all kinetic terms of the gravitational field vanish:
This is easily seen by splitting the kinetic term as follows:
\be
\pi^{ij}\dd_t g_{ij}=(\pi^T)^{ij}\dd_t g^T_{ij}+\pi\dd_t e^\phi
\label{kineticterm}
\ee
where $\pi\equiv \pi^l_l$ and a traceless tensor $P^T_{ij}$ is defined by
\be
P^T_{ij}=P_{ij}-\frac{1}{2}P^l_l\de_{ij}
\label{tracelesstensor}
\ee
So because $g^T_{ij}=0$ and $\pi=0$ we see that this kinetic term vanishes.
Furthermore the term: $NH+N_iH^i$, which also contains matter contributions,
must
vanish because these are precisely the
constraints that we solve.
\footnote{More precisely, the set of first class constraints $\{H,H^i\}$
becomes
after imposing the gauge conditions (\ref{conformalmetric}) and (\ref{piis0}) a
twice as large set of second class constraints which can be set equal to 0
strongly
on the constraint surface.}
The reduced Lagrangian is thus given by the kinetic term of the matter fields
and the surface
term (\ref{surfaceterm}) which now only depends on the matter
phase space variables because we (formally) solved the gravitational fields
in terms of the matter variables. The surface term thus serves as a hamiltonian
for the matter degrees of freedom. We will see later that it represents
actually
the total deficit angle of the universe (see \cite{Henneaux}).
As an example, we will work out the solution found in \cite{beginarticle}
for a massive, spinning particle at the origin, in our gauge. The solution for
the metric will
become degenerate near the particle. We are however not interested in that
region since it is
exactly the place where closed time-like curves may occur. At infinity this
solution is however
very interesting as we expect that any configuration of particles can be
described asymptotically
by an effective center of mass particle with mass M equal to the total energy
contained in the
universe and spin S equal to the total angular momentum of the universe. The
solution therefore
will provide the asymptotic behaviour of the gravitational fields.

If we use flat coordinates $u^a$ the matching conditions for this particle
(sitting at rest  in the origin) will be \cite{beginarticle}:
\ba
\tilde{u}^a&=& R^a_{~b}u^b+q^a \label{transformationu}\\
R^a_{~b}&=& \left( \begin{array}{ccc}
          1&0&0\\
          0&\cos 2\pi\al&\sin 2\pi\al\\
          0&-\sin 2\pi\al&\cos 2\pi\al  \\
             \end{array} \right) \label{rotation} \\
q^a&=& \left( \begin{array}{c}
               2\pi A \\
               0\\
               0      \end{array} \right) \label{translation}
\ea
Here $\tilde{u}^a$ is the value of $u^a$ after an anti-clockwise rotation about
the particle.
Where we have defined: $\al=4GM$ and $A=4GS$. For a general choice of origin
and a moving
particle the identification will be an element of the Poincar\'e group:
\be
\tilde{u}^a = (BRB^{-1})^a_{~b}u^b+q^a\label{Lorentz}
\ee

Here $B^a_{~b}$ is a boost matrix with arbitrary rapidity $(\eta)$ and in an
arbitrary direction,
and $q^a$ is now a general translation (not necessarily of the form
(\ref{translation})).
The line-element for the massive spinning particle sitting in the
origin (not moving), found in \cite{beginarticle}, is given by:
\be
ds^2=-(dt+Ad\theta)^2+\frac{1}{r^{2\al}}(dr^2+r^2d\ta^2) \label{line-element}
\ee
This is clearly not in the gauge we have chosen and to transform this into our
gauge we have
to write it as:
\be
ds^2=-dt^2-2Adtd\ta(\rho,\omega)+e^\phi(d\rho^2+\rho^2d\omega^2)
\label{line-element2}
\ee
where the difference is in the $A^2d\theta^2$ term.
We took here also $g_{oo}=-1$ because the system is stationary and
$g_{oo}=\dd_t u^a\dd_t u^b\eta_{ab}$.
The map from the $(t,r,\theta)$-coordinate system to the
$(t,\rho,\omega)$-coordinate system is provided by:
\ba
\omega&=&\ta \label{w-teta} \\
\ln\frac{\rho}{\rho_0}&=&\frac{1}{1-\al}\ach(\frac{r^{1-\al}} {A})|^r_{r_0}
\label{rho-r}
\ea

If we choose ${r_0}^{1-\al}=A$ and $\rho_0=\rho(r_0)=1$ and use the complex
coordinates: $\rho=z\bar{z},\omega=\frac{1}{2i}\ln(\frac{z}{\bar{z}})$
we find for the line element in our gauge:
\be
ds^2=-dt^2+\frac{iA}{z}dtdz-\frac{iA}{\bar{z}}d\bar{z}dt+\frac{A^2}{4z\bar{z}}
[\frac{(z\bz)^{1-\al}}{A^2}-2+A^2(z\bz)^{\al-1}]dzd\bz \label{line-element3}
\ee
 So this will be the asymptotic behaviour of any metric describing N moving
point-particles
with total energy $\frac{\al}{4G}$ and total angular momentum $\frac{A}{4G}$.
{}From here one finds that the coordinates $u^a$ are given by:
\ba
 u^0&=& t+\frac{A}{2i}\ln z-\frac{A}{2i}\ln\bar{z}\label{u0=}\\
 u &=&\frac{1}{\sqrt{2}(1-\al)}[z^{1-\al}+\frac{A^2}{\bz^{1-\al}}]\label{u=}\\
\bar{u}&=&\frac{1}{\sqrt{2}(1-\al)}[\bz^{1-\al}+\frac{A^2}{z^{1-\al}}]\label{ubar=}
\ea
We see that both the $u^0$ and the $u$ and $\bar{u}$ are multivalued. Note that
the log terms
in $u^0$ are needed to generate the time jump of $8\pi GS$.
The dreibein fields are given by:
\ba
\dd_0 u^a=e_0^a &=& (1,0,0) \label{e0=}\\
\dd_z u^a=\tet&=&(\frac{A}{2iz},
\frac{1}{\sqrt{2}z^\al},-\frac{A^2 z^{\al-2}}{\sqrt{2}})
\label{ez=}\\
\dd_{\bz} u^a=\tec&=& (-\frac{A}{2i\bz},-\frac{A^2
\bz^{\al-2}}{\sqrt{2}},\frac{1}{\sqrt{2}\bz^\al})\label{ezbar=}
\ea
one can easily check that these dreibein fields fullfill our gauge conditions
(\ref{ez},\ref{ebarz},\ref{nullez},\ref{nullebarz}). These coordinates $u^a$
and the
3 dreibeins are only determined up to a global Lorentztransformation, since
$\Lambda^a_{~b} u^b$
would produce the exact
same line element (\ref{line-element3}). Actually we can choose our global
Lorentzframe by the
condition that the dreibeins have this asymptotic form.
This immediately implies that the constant function $C^a(t)$ in (\ref{e0}) is
determined to
be $\de^a_0$. In general however we will allow for more general global
Lorentzframes
(i.e. frames in which the center of mass moves). It will be this this global
frame that determines the function $C^a(t)$.

\section{the Riemann-Hilbert problem}
The major advantage of our gauge choice is that the dreibein field splits into
a
holomorphic and a anti-holomorphic piece (and $e_0^{~a}$). In effect we only
have to solve the $\tet$-field because the $\tec$ is then given by:
\be
\tec=F^a_{~b}(e^{~b}_z)^*~~~~~F=\left( \begin{array}{ccc}
                                         1&0&0\\
                                         0&0&1\\
                                         0&1&0
                                         \end{array} \right)\label{e=e*}
\ee
Then, in the Lorentzframe where the effective center of mass is not moving, the
$\teo$-field is given by the the expression:
\be
\teo=\de^a_0+\dd_t\int dz~\tet +\dd_t\int d\bz~\tec\label{e02}
\ee
So we only have to solve for the null-vector $\tet$, living on the complex
plane.
We will not try to solve the Einstein-equations explicitly but we will use the
monodromy properties of the dreibein. Now what happens if we parallel transport
the dreibein $\tet$ over a loop $\gamma_n$ around a point particle at position
$a_n$. First we notice that we can deform the loop $\gamma_n$ as long as
we don't pass through a particle. This means that the tranformation will
only differ on different homotopy classes of loops. To be more precise, we
are interested in the transformations from the first homotopy class of loops
on a punctured Riemann surface, to the group SO(2,1):
\ba
&& \pi_1(C-\{a_1,...a_N,\infty\};z_o)\rightarrow \mbox{SO(2,1)}
\label{homotopy}\\
&& [\gamma]_n \rightarrow M_n \nn
\ea
We would like to view the coordinates $u^a(z)$
\footnote{The vector $u^a(z)$ is the holomorphic part of the vector
$u^a(z,\bz)=u^a(z)+u^a(\bz)$ (see section 2).}
as multivalued functions on the
complex plane. We therefore need a cut in the surface where the functions
$u^a(z)$
transform. We attach to each particle a string over which the $u^a(z)$
transform as
a Poincar\'e vector:
\be
u^a\ra(BRB^{-1}\cdot u)^a +l^a \label{transformationu2}
\ee
At this point we should warn the reader that the translation vector $l^a$ is
not the same as the translation vector $q^a$ in (\ref{Lorentz}).
This is due to the fact that we consider here only the holomorphic part of
$u^a$.
For the anti-holomorphic part we have a similar transformation with a
translation
vector $(l^*)^a$. The correspondence is thus given by $q^a=l^a+(l^*)^a$,
which is a real vector. The transformation law (\ref{transformationu2}) implies
that the $\tet=\dd u^a$ transform under the homogeneous part of the the
Poincar\'e group; the Lorentzgroup. So if we parallel transport the
dreibein along a closed curve (loop) $\gamma_n$ we find:
\be
\tet\stackrel{[\gamma]_n}{\rightarrow}
(B_nR_nB^{-1}_n\cdot e_z)^a \label{transformatione}
\ee
As the particles evolve we don't allow them to hit a string, so one cannot
avoid that
the strings might get 'knotted' in a very complicated way.
\begin{figure}[t]
\centerline{\psfig{figure=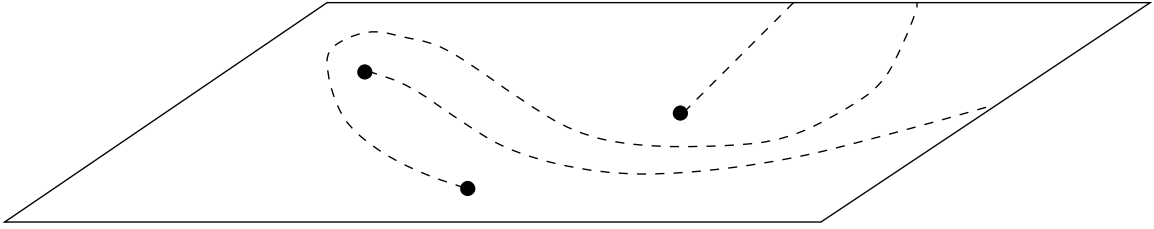,angle=-90,height=3cm}}
\caption{Particles with strings attached moving on the complex plane.}
\label{particleswithstrings}
\end{figure}
On the other hand, if we have an expression for $\tet$ we can reconstruct
$u^a(z)$
by integrating once:
\[ u^a(z)=\int dz~\tet \]
It is now easy to see that the multivaluedness of $u^a(z)$ is indeed of the
form
(\ref{transformationu2}). If we integrate around a singularity at $a_i$ we
find:
\ba
\tilde{u}^a(z_o)&=&\int_{z_o}^{a_k}dz~\tet+\oint_{a_k}dz~\tet
+M_k\int_{a_k}^{z_o}dz~\tet\\
&=&(1-M_k)A^a_k+\oint_{a_k}dz~\tet
\ea
where $A^a_i\equiv\int_{z_o}^{a_i}dz~\tet$ is the position of the particle in
flat
coordinates, and $M_k=B_kR_kB_k^{-1}$ (the monodromy matrix).
Because $u(z_o)=0$ is the origin of the flat coordinate system, the
$\tilde{u}^a$
represent the shift of the origin and as such is equal to the vector $l^a$ of
(\ref{transformationu2}).
If we integrate around 2 particles we find:
\be
\tilde{u}^a(z_o)=(1-M_1)A^a_1+M_1(1-M_2)A^a_2+\oint_{a_1}dz~\tet+M^a_{1~b}
\oint_{a_2}dz~e^b(z)
\ee
{}From this expression, together with the antiholomorphic part, we can derive
the relative
angular momentum of the 2 particles. It is defined in the  center of mass frame
of the 2 particles by:
\be
L=[B^{-1}_{\rm c.o.m.}\cdot(l+l^*)]^0
\ee
This is precisely the same expression as in \cite{beginarticle} only if we
demand:
\be
\lim_{R\ra 0}\oint_{a_i}dz~\tet=0
\label{conditionR}
\ee
where we take the radius $R$ of intgration to go to zero. To produce the
multivaluedness
of the $\tet$ we expect, in the frame where the particle is boosted to rest,
the following local behaviour near
the location of a particle $a_i$:
\be
(B^{-1}_i e_z)^a\stackrel{z\ra a_i}{\sim}~z^{\lambda^a_i}H_i^a(z)
\ee
The $\lambda^a_i$ (called local exponents) are real numbers and the $H_i^a(z)$
are holomorphic
functions near $a_i$. The condition (\ref{conditionR}) now implies that:
\be \lambda^a_i>-1  \label{lala}\ee
In fact we would like that for small $\al_i=4Gm_i$, the results do not differ
much from the
flat metric, i.e. the metric $g_{\mu\nu}$ is just  slightly perturbed from the
flat metric $\eta_{\mu\nu}$.
This fact and the fact that $(B^{-1}e_z)^a$ must produce a cut in the $z$-plane
across which
a vector must rotate over an angle $2\pi\al_i$ determines the local exponents
to be:
\be
\lambda^a_i=\left(\begin{array}{c}
                      0\\ -\al_i\\ \al_i \end{array}\right) \ee
See also \cite{Bellini}. Now we must take $\al_i\in~(0,1)$ in order to fullfill
relation
(\ref{lala}). Also if we would exceed this limit the total mass would become
too large and
the universe would close.\\
Note that if we would transform to a different global Lorentzframe, using a
Lorentztransformation $\Lambda$, the effect on the monodromy would be
\be
M_n\ra\Lambda^{-1}M_n\Lambda \label{frame}
\ee
This is just a conjugation that doesn't change the structure constants of the
group (SO(2,1)).
So the abstract group remains the same.  Finally it is important to remark that
if we follow a
trajectory around all particles in the positive direction (anti-clockwise),
this is equivalent
to follow a trajectory around infinity in the negative direction, i.e.:
\be
M_{tot}=B_{\infty}R^{-1}_{\infty}B^{-1}_{\infty}=
B_NR_{N}B^{-1}_N.....B_1R_1B_1^{-1}\label{Mtotaal}
\ee
We will treat infinity as a singular point on the Riemann sphere, just as the
particle positions.
However because we demand that our universe is open the nature of this
singularity will be slightly
different from the particle singularities. We are now ready to reformulate our
problem of solving the
$\tet$ as an old mathematical problem, \underline{the Riemann-Hilbert
problem:}\\
Given the following transformations for parallel transporting a vector around a
closed loop:
\ba
&&\tet\stackrel{[\gamma]_n}{\rightarrow} M_{n~~b}^{~a}e_z^{~b}
\label{transformatione2}\\
&&[\gamma]_n \in \pi_1(C-\{a_1,...a_N,\infty\};z_o)\nn\\
&&M_n\in SO(2,1)\nn
\ea
Find the multivalued, holomorphic (=meromorphic) functions $\tet$ that
transform in this way,
with initial value $e^{~a}_z(z_o)=\stackrel{o}{e}_z^{~a}$.\\
It is actually more convenient to work in the locally isomorphic group SU(1,1).
Denote $\zeta^\al(z)$ to be a spinor, transforming under SU(1,1) and
$\zeta^{*\al}$
to be a spinor transforming under the complex conjugate representation of
SU(1,1).
(Note that $\zeta^{*\al}$ is not the complex conjugate of $\zeta^\al$.) If the
components of $\zeta^\al$ are given by
\be
 \zeta^\al= \left(\begin{array}{c}
                    a\\b \end{array}\right)
\ee
then the spinor transforming under the complex conjugate representation is:
\be
 \zeta^{*\al}=\bt\left(\begin{array}{c}
                    b\\a \end{array}\right)
\label{zeta}
\ee
where $\bt$ is an arbitrary real constant.
The vector $\tet$ transforming under SO(2,1) is now constructed as follows:
\be
e^a=\bt\left( \begin{array}{c}
                   2ab\\
                    -i\sqrt{2} a^2\\
                     i\sqrt{2} b^2
                    \end{array} \right)\label{e-zeta}
\ee
Note that we automatically construct a null-vector $\tet$ in this way.
\footnote{The fact that we can express the components of $\zeta^*$ in
the components of $\zeta$ is a consequence of the fact that for SU(1,1)
the conjugate representation is not independent but can be written as:
\be
M^*=\tau^{-1}M\tau~~~~~~\tau=\left( \begin{array}{cc}
                         0&1\\
                       1&0 \end{array}\right) \label{geconjungeerdemonodromy}
\ee
}
In the following we will always use this spinor representation SU(1,1) of the
Lorentzgroup in 2+1 dimensions. The (modified) Riemann-Hilbert problem we have
to solve is thus:
\ba
&&\zeta^\al\stackrel{[\gamma]_n}{\ra}M^\al_{n~\bt}\zeta^\bt
\label{RHprob2}\\
&&[\gamma]_n\in \pi_1(C-\{a_1...a_N,\infty\},z_o)\nn\\
&&M_n\in SU(1,1)\nn
\ea
The Riemann-Hilbert problem has a long and interesting history. Riemann was the
first (1857) to work with a system of functions satisfying a linear
differential
equation with a {\em given} monodromy group (however, he couldn't prove that
this
system of functions exists). In 1912 Hilbert added this problem to his list and
since then many people worked on the problem (and solved it; that is, proved
that
there is a system of functions with a given monodromy group). People who worked
on
the problem were: Plemelj \cite{Plemelj}, Birkhoff, R\"ohrl (who proved the
statement
for an arbitrary Riemann surface) and Lappo-Danilevsky. Another important
breakthrough
was the reduction of the Riemann problem to a system of Schlesinger equations.
For a
review on this see \cite{Chudnovsky,Sato}. We will follow in this paper two
important
ways to solve the Riemann-Hilbert problem. First we will use a second order
Fuchsian
differential equation to attack the problem and solve the simple case of 2
particles.
Then we will use the approach of a system of first  order differential
equations and
the Schlesinger deformation equations to write down a general (formal) solution
first
found by Miwa, Sato and Jimbo \cite{Sato}.
\subsection{The Fuchsian differential equation}
It was already recognized by Riemann that differential equations with regular
singularities have a monodromy group. If one would take the d independent
solutions to the d'th order linear differential equation with N singular points
(infinity is not included in the number N, but it is singular) and move them
around a singular point they would transform into a linear combination of them.
The general form of the differential equation with only regular singular points
(i.e. Fuchsian singularities, see below) was found by Fuchs and reads:
\be
\dd^{(d)}\zeta+\frac{P_{(N-1)}(z)}{Q(z)}\dd^{(d-1)}\zeta+
\frac{P_{2(N-1)}}{Q^2(z)}\dd^{(d-2)}\zeta+....=0 \label{Fuchsdv}
\ee
$P_{l(N-1)}$ is a polynomial of order $l(N-1)$ and
$Q=\prod^{N}_{k=1}(z-a_k).$\\
As we will use the 2 dimensional representation of the monodromy group SU(1,1)
we can take d=2.
If we had chosen to work with SO(2,1) we would have to take d=3.
We want that the 2 independent solutions $\zeta^1$ and $\zeta^2$ to this
equation are "regular"
singular near the
points $a_k$. This implies that by definition they must have the following
behaviour near $a_k$:
\be
\zeta^\al=\stackrel{z\ra
a_k}{\sim}(z-a_k)^{\lambda^\al_k}[b^\al_k+c_k^\al(z-a_k)+...]
\label{localbehaviour}
\ee
This local behaviour is defined in a local basis where the monodromy is
diagonal. In general the
solution near a singular point is a linear combination
of the 2 solutions ($\al=1,2$) given above (see (\ref{P})).
The coeficients $\lambda^\al_k$ (called the local exponents) must obey the
Fuchs relation:
\be
\sum_{\al=1}^d\sum_{k=1}^N \lambda^\al_k +\sum_{\al=1}^d
\lambda^\al_\infty=\frac{d(d-1)}{2}(N-1)
\label{somexponents}
\ee
The first  non-trivial case appears with 2 singularities (and infinity).
This is also a special case because as we will explicitely
show, the local exponents completely determine the Fuchsian differential
equation and its
solutions.
 With N singular points there will be (N-2) undetermined parameters (free
parameters) in
the differential equation.
 They must be determined by matching the monodromy matrices $M_n$ with the
general
monodromy of the solutions to that equation. If we have more than 2
singularities a
complication is that apparent singularities may occur. (at most N-2
\cite{Chudnovsky}).
These are singular points in the differential equation
but not in the solution. They will however generate zero's in the $\zeta^\al$.
At a zero the local exponents are possitive integers, and they are counted in
the sum
(\ref{somexponents}). This is reason why for more than 2 particles the
method of finding a d'th order Fuchsian differential equation becomes very
difficult and we use another method.
Another remark is that the Riemann-Hilbert problem is not uniquely solvable if
we only
give the monodromy matrices. Let us define $\vec{\phi}(z)$ to be the d
independent
solutions to a definit Fuchsian
differental equation. For a given monodromy there will exist d such sets of
independent
functions $\vec{\phi}_i(z)$, which are solutions to d Fuchsian differential
equations with
the same monodromy group but with different
exponents at the singularities.
These d sets are rationally independent
with respect to eachother. I.e. a new set can always be written as:
\be
\vec{f}(z)=\sum_{i=1}^{d}q_i(z)\cdot\vec{\phi}_i(z)
\label{rationalcombinations}
\ee
and $q_i(z)$ are rational functions. (rational functions will not change the
monodromy of the solutions). A basic result is \cite{Blok}:\\
Only if we specify the local exponents $\lambda^\al_k$ exactly (so not only up
to an integer that would not change the monodromy, but also the integer part)
then
we can find a unique solution to the Riemann-Hilbert problem. As it turns out,
the
local exponents at $a_k$ are proportional to the masses of the particles and
the local
exponent at infinity to the total mass. As we saw in section 3, physics demands
that we
choose the masses in the range $(0,1)$. This   determines the integer part of
the exponents
and makes the problem uniquely solvable.
This method of finding a Fuchsian differential equation that has the right
monodromy is
particularly suited for the 2 particle case. We will use a different strategy
for the
general case. Below we will work out the 2 particle solution.

\subsection{The 2 particle solution}
To solve the 2 particle case one has to notice that a second order Fuchsian
differential equation with 3 regular singularities, one of which is located
at infinity, can always be written in the following form:
\be
\dd^2\zeta+\sum^2_{k=1}\frac{1-\la^1_k-\la^2_k}{z-a_k}\dd\zeta+
\{\sum^2_{k=1}\la^1_k\la^2_k\frac{(a_k-a_{k+1})}{z-a_k}+
\la^1_\infty\la^2_\infty\}\cdot\frac{\zeta}{\prod_{i=1}^{2}(z-a_i)}=0
\label{Papperitz}
\ee
where we have defined $a_k$ cyclic: $a_3=a_1$ etc. The general
solution can be written symbolically using the Riemann-scheme:
\be
\zeta=P\left\{ \begin{array}{cccc}
               a_1 & a_2 & \infty &  \\
               \la^1_1&\la^1_2&\la^1_\infty&;z\\
               \la^2_1&\la^2_2&\la^2_\infty&
                \end{array} \right\}
\label{Riemann-scheme1}
\ee
{}From this notation one can find in the upper row the singular points and in
the
second and third row the local exponents.
Remember that these local exponents have to satisfy the Fuchs-relation:
\be
\sum_{k,\al=1}^2\la^\al_k+\sum_{\al=1}^2\la^\al_\infty=1
\label{Fuchs}
\ee
Because we still had conformal freedom left in our gauge we can bring the
points
$\{a_k\}$ to respectively $\{0,1\}$ using the following conformal
transformation:
\be
w=\frac{(z-a_1)}{(a_2-a_1)}
\label{fractransf}
\ee
We would like to transform this equation to the hypergeometric equation. This
is possible by a
transformation to a new variable $\xi(w)$, defined by:
\be
\zeta(w)=w^{\la^1_1}(1-w)^{\la^1_2}\xi(w)
\label{trans1}
\ee
The Riemann-scheme now looks like:
\be
\xi=P\left\{ \begin{array}{cccc}
               0 & 1 & \infty &  \\
               0 & 0 & \la^1_1+\la^1_2+\la^1_\infty&;w\\
               \la^2_1-\la^1_1&\la^2_2-\la^1_2&\la^1_1+\la^1_2+\la^2_\infty&
                \end{array} \right\}
\label{Riemann-scheme2}
\ee
And this is precisely the Riemann-scheme for the hypergeometric function
$F(a,b,c;w)$ with coeficients:
\begin{eqnarray}
a&=&\la^1_1+\la^1_2+\la^1_\infty \nn\\
b&=&\la^1_1+\la^1_2+\la^2_\infty\nn\\
c&=&1+\la^1_1-\la^2_1
\label{coefficients}
\ea
There exist 2 independent solutions to the hypergeometric equation near $w=0$
(the series converges in the circle $|w|<1$):
\ba
\xi^1&=&F(a,b,c;w)\\
\xi^2&=&w^{1-c}F(1+a-c,b-c+1,2-c;w)
\ea
And also 2 rationally independent ones which one obtains by taking
$\tilde{\la}^1_\infty=\la^1_\infty+1$ and
$\tilde{\la}^2_\infty=\la^2_\infty-1$. We will
however use the first ones, because they provide us with the right asymptotic
behaviour.
This means that the solution for $\zeta^\al$ around $w=0$ becomes (using some
basic properties of F):
\ba
\zeta^1&=&w^{\la^1_1}(1-w)^{\la^1_2}F(a,b,c;w)\\
\zeta^2&=&w^{\la^1_2}(1-w)^{\la^2_2}F(1-a,1-b,2-c;w)
\label{zeta2}
\ea
We will call this solution $\zeta^\al=(P(\la^1_1),P(\la^2_1))$. The argument of
P determines the local exponent of that solution. If we want to find the
solutions
outside the circle $|w|=1$ we have to take the analytic continuation of this
solution to the rest of the complex plane. Of course this is all well known
for the hypergeometric equation. Denoting the 2 independent solutions to the
Papperitz equations near $w=1$ and $w=\infty$ in the way as above we find that
the analytic continuation looks like:
\ba
P(\la^1_1)&=&A^1_2P(\la^1_2)+A^2_2P(\la^2_2)\nn\\
P(\la^2_1)&=&B^1_2P(\la^1_2)+B^2_2P(\la^2_2)\nn\\
P(\la^1_1)&=&A^1_\infty P(\la^1_\infty)+A^2_\infty P(\la^2_\infty)\nn\\
P(\la^2_1)&=&B^1_\infty P(\la^1_\infty)+B^2_\infty P(\la^2_\infty)
\label{P}
\ea
The coefficients can for instance be found in \cite{complex}. Using the local
expansion (\ref{localbehaviour}) we can simply see what happens if we move
around the
location $a_k=\{0,1,\infty\}$. The $P(\la^\al_k)$  will change as follows:
\be
P(\la^\al_k)\ra e^{2\pi i\la^\al_k}  P(\la^\al_k)
\label{trafoP}
\ee
we can actually derive the monodromy explicitely (\cite{complex}).
In the following we will give these monodromy matrices and match them with the
monodromy we impose for a particle.
First we define the following basis:
\ba
u&=&R~\Lambda ~P(\la^1_1)\\
v&=&S~P(\la^2_1)
\label{basis}
\ea
Here R and S are functions of the parameters of the system only and $\Lambda$
is:
\ba
\Lambda&=&\frac{\Gamma(1+\la^2_1-\la^1_1)\Gamma(\la^1_1+\la^2_2+\la^1_\infty)\Gamma(\la^1_1+\la^2_2+\la^2_\infty)}
{\Gamma(1+\la^1_1-\la^2_1)\Gamma(\la^2_1+\la^2_2+\la^1_\infty)\Gamma(\la^2_1+\la^2_2+\la^2_\infty)}\nn\\
&-&\frac{\Gamma(1+\la^2_1-\la^1_1)\Gamma(\la^1_1+\la^1_2+\la^1_\infty)\Gamma(\la^1_1+\la^1_2+\la^2_\infty)}
{\Gamma(1+\la^1_1-\la^2_1)\Gamma(\la^2_1+\la^1_2+\la^1_\infty)\Gamma(\la^2_1+\la^1_2+\la^2_\infty)}
\label{gamma}
\ea
Now the monodromy for moving around $w=0$ is:
\be
M_1=\left( \begin{array}{cc}
            e^{2\pi i\la^1_1}&0\\
              0 & e^{2\pi i\la^2_1}
             \end{array}\right)
\label{mon1}
\ee
And the monodromy for moving around $w=1$ is:
\be
M_2=\left( \begin{array}{cc}
            \frac{\mu e^{2\pi i\la^1_2}-e^{2\pi i\la^2_2}}{\mu-1}
&\frac{R}{S}(e^{2\pi i\la^2_2}-e^{2\pi i\la^1_2})\\
             \mu\frac{S}{R}\frac{(e^{2\pi i\la^1_2}-e^{2\pi i\la^2_2})}
{(\mu-1)^2}&\frac{\mu e^{2\pi i\la^2_2}-e^{2\pi i\la^1_2})}{\mu-1}
             \end{array} \right)
\label{mon2}
\ee
\be
\mu=\frac{\sin\pi(\la^1_1+\la^2_2+\la^2_\infty)\sin\pi(\la^2_1+\la^1_2+\la^2_\infty)}
{\sin\pi(\la^2_1+\la^2_2+\la^2_\infty)\sin\pi(\la^1_1+\la^1_2+\la^2_\infty)}
\label{mu}
\ee
This monodromy has to be matched with the desired SU(1,1) monodromy. We will
fix particle 1
(at $w=0$) and allow particle 2 (at $w=1$) to move
in an arbitrary direction with an arbitrary velocity (see Fig. 3).
\begin{figure}[t]
\centerline{\psfig{figure=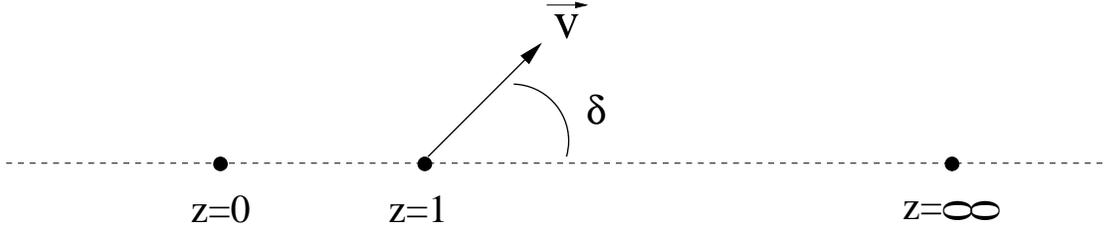,angle=-90,height=3cm}}
\caption{Two particles moving on the complex plane}
\label{2particles}
\end{figure}
The monodromy for particle 1 is then:
\be
R_1=\left( \begin{array}{cc}
            e^{-i\pi\al_1} & 0\\
            0 & e^{i\pi\al_1}
            \end{array} \right)~~~~~~~~~\al_i=4Gm_i
\label{rot1}
\ee
The second particle has a more general monodromy of the form:
\be
B_2R_2B_2^{-1}=\left( \begin{array}{cc}
                \cos(\pi\al_2)-i\sin(\pi\al_2)\cosh\eta &
  -\sin(\pi\al_2)\sinh\eta e^{i\delta}\\
  -\sin(\pi\al_2)\sinh\eta e^{-i\delta}&
  \cos(\pi\al_2)+i\sin(\pi\al_2)\cosh\eta \end{array} \right)
\label{BRB}
\ee
The monodromy around $w=\infty$ is now fixed by the relation:
\be
M_1M_2M_\infty=I
\label{mon3}
\ee
Matching the monodromies gives:
\ba
2\la^1_1&=&-\al_1~~~~\mbox{(mod $n_1$)}\nn\\
2\la^2_1&=&\al_1~~~\mbox{(mod $n_2$)}\nn\\
2\la^1_2&=&-\al_2~~~~\mbox{(mod $n_3$)}\nn\\
2\la^2_2&=&\al_2~~~\mbox{(mod $n_4$)}
\label{lambda=alpha}\\
\mu&=&\coth^2(\frac{1}{2}\eta)
\label{mu=coth}\\
\frac{R}{S}&=&\frac{i}{2}\sinh\eta e^{i\de}
\label{R/S}
\ea
We take here (mod $n_i$) $n_i\in Z$, instead of (mod $2n_i$). This has to do
with the fact that SU(1,1) is the covering group of SO(2,1). In this case this
implies both the transformations $\pm S\in$ SU(1,1) correspond to one element
in SO(2,1). For reasons explained before (section 3) we will take $n_i=0$ (See
also \cite{Bellini}).
We can simplify relation (\ref{mu=coth}) to:
\be
\cos(2\pi\la^2_\infty)=\cos(2\pi\la^1_1)\cos(2\pi\la^1_2)-\cosh\eta\sin(2\pi\la^1_1)\sin(2\pi\la^1_2)
\label{rel-m-eta}
\ee
This is the precisely the same relation found in \cite{beginarticle}. We will
therefore interpret
$4\pi\la^2_\infty$ as the total deficit angle of the universe which means that
$\al_{tot}\equiv 2\la^2_\infty$
must be in the range (0,1). Next we can use the Fuchs relation (\ref{Fuchs}) to
find $\la^1_\infty$:
\be
\la^1_\infty=1-\la^2_\infty
\label{lambda1,2}
\ee
The complex conjugate representation can be retrieved by using (\ref{zeta}).
Now we can write down the full solution to the monodromy problem:
\be
\zeta^\al(w)=\bt\left( \begin{array}{l}
                 \frac{i}{2}\sinh\eta e^{i\de}\Lambda
P(-\al_1,\al_2,\al_{tot};w)\\
                  P(\al_1,\al_2,\al_{tot};w)
                   \end{array}\right)
\label{oplossing-zeta}
\ee
with:
\be
P(\al_1,\al_2,\al_{tot};w)=w^{\frac{\al_1}{2}}(1-w)^{\frac{\al_2}{2}}
F(1+\frac{\al_1}{2}+\frac{\al_2}{2}-\frac{\al_{tot}}{2},\frac{\al_1}{2}+\frac{\al_2}{2}+\frac{\al_{tot}}{2},1+\al_1;w)
\label{P=}
\ee
Using the formula (\ref{e-zeta}) we can write down the dreibein:
\be
\tet=\bt\left( \begin{array}{l}
             i\sinh\eta e^{i\de}\Lambda P(\al_1)P(-\al_1)  \\
             \frac{i}{2\sqrt{2}}\sinh^2\eta e^{2i\de}\Lambda^2 P^2(-\al_1)\\
              i\sqrt{2} P^2(\al_1)
           \end{array} \right)
\label{oplossing-e}
\ee
It is now interesting to check the asymptotic behaviour found in formula
(\ref{oplossing-zeta}).
Since we know the local exponents at infinity
we can write down the following expansion for $P(\la^\al_\infty)$:
\be
P(\la^\al_\infty)=(\frac{1}{w})^{\la^\al_\infty}[b^\al_\infty+\frac{c^\al_\infty}{w}+.....]
\label{expansionP}
\ee
with
\ba
\la^1_\infty&=&\frac{\al_{tot}}{2}\label{lambda-altot1}\\
\la^2_\infty&=&1-\frac{\al_{tot}}{2}\label{lambda-altot2}
\ea
Using (\ref{e-zeta}) we find the asymptotic behaviour for $\tet$:
\be
\tet\stackrel{w\ra\infty}{\sim}\left( \begin{array}{l}
                                       w^{-1}\\
                                       w^{-\al_{tot}}\\
                                       w^{\al_{tot}-2}
  \end{array} \right) [B^a_\infty+\frac{C^a_\infty}{w}+.....]
\label{expansion-e}
\ee
It is important to realize that the global Lorentzframe we use to calculate the
asymptotic behaviour is different from the one we used to calculate the
$P(\la^\al_1)$.
In this frame particle 1 is not moving and as a result $M_1$ is diagonal in the
basis $P(\la^\al_1)$. In the frame we used to derive the asymptotic behaviour
the
monodromy around infinity is diagonal which means that the effective center of
mass
is not moving. The relation between the 2 frames is given by the coefficients
$A^\al_\infty$ and $B^\al_\infty$ of (\ref{P}). We see that the asymptotic
behaviour
is the same as in (\ref{e0=},\ref{ez=},\ref{ezbar=}). One could also try to
find the
total angular momentum by matching the first coefficient in the expansion with
(\ref{e0=},\ref{ez=},\ref{ezbar=}):
\be
B^a_\infty=(\frac{A}{2i},\frac{1}{\sqrt{2}},-\frac{A^2}{\sqrt{2}})
\label{matchinfty}
\ee
\subsection{the N-body case}
To treat the N-body case we will use a completely different approach, first
followed by Schlesinger. Every monodromy problem will correspond to a
first order matrix differential equation. First we write
$\zeta^\al(z)=Y^\al_{~~\bt}\zeta^\bt(z_o)$ where $Y(z)$
will be a fundamental matrix of solutions to the differential equation. The
$Y(z)$ has the following properties:
\begin{enumerate}
\item[i)] $Y(z_o)=I$
\item[ii)] $Y(z)$ is holomorphic in $(C-\{a_1...a_N,\infty\})$
\item[iii)] $Y(z)$ has the following short distance behaviour near a singular
point:
\end{enumerate}
\be
Y(z)\stackrel{z\ra a_i}{\sim}(z-a_i)^{L_i}\hat{Y}_i(z)
\label{expY}
\ee
where $M_i=e^{2\pi i L_i}$ and $M_i$ are the monodromy matrices. It is also
important that the
matrix $\hat{Y}_i(z)$ is holomorphic and invertable at $z=a_i$. So:
\be
\hat{Y}_i(z)\stackrel{z\ra a_i}{\sim}(Y^0(a_i)+Y^1(a_i)(z-a_i)+....)
\label{hatY}
\ee
with $\det|Y^0(a_i)|\neq0$. \\
To derive the form of the differential equation we first consider the object
$Y^{-1}(\dd Y)$.
This is also a single valued expression (the monodromy cancels out) and
holomorphic at $(C-\{a_1...a_N,\infty\})$.
By  using the local behaviour near the singularity $a_i$ we can find an
expression near the location $a_i$:
\be
Y^{-1}(\dd Y)\stackrel{z\ra a_i}{\sim}
\hat{Y}^{-1}_i(z)\frac{L_i}{z-a_i}\hat{Y}_i(z)+\mbox{holomorphic at $z=a_i$}
\label{expY-1dY}
\ee
Hence it must be of the form:
\be
Y^{-1}(\dd Y)=\sum_{i=1}^{N}\frac{A_i}{z-a_i}
\label{diffeq}
\ee
with
\be
A_i=\mbox{Res}_{z\ra a_i}Y^{-1}(\dd Y)=(Y^0_i)^{-1}L_iY^{0}_i
\label{A=}
\ee
The behaviour of this equation at infinity can be studied by
transforming to a new coordinate $w=\frac{1}{z}$. We find that that:
\be
A_\infty=-\sum_{i=1}^{N}A_i
\label{A_infty=}
\ee
To prove that the solution to the differential equation:
\be
\dd Y=\sum_{i=1}^{N}Y\frac{A_i}{z-a_i}
\label{dv}
\ee
is unique we consider 2 solutions $Y_1$ and $Y_2$ and study the function $Y_1
Y_2^{-1}$.
Using the conditions ii) and iii) we see that this function is single valued
and
holomorphic everywhere on the Riemann sphere. This implies that it must be
constant.
Using i) we find that it must be actually $I$ and we have proved the statement:
$Y_1=Y_2$,
which in turn proves that the solution is unique. The proof that there actually
exists a
solution is much more involved and was only esthablished for the first time in
its full
generality by R\"ohrl. We will state the result obtained by Lappo-Danilevsky.
He obtained a formal solution in terms of expansions in hyperlogarithms. See
\cite{Chudnovsky}
for a review. He proved that if the $L_i$ were sufficiently small to ensure the
convergence of
this series, then the Riemann problem has d rationally independent solutions.
Every other
solution is given as a rational combination of these solutions. Remember that
if we multiply
$Y$ with a rational matrix $R(z)$ the monodromy does not change. Even when no
convergence was
guaranteed the solution could still be interpreted as analytic continuations.
If we furthermore specify the local exponents exactly, that is specify the
local behaviour of
the spinor $\zeta^\al$ exactly (in a canonical basis), then we can find a
unique solution for
$\zeta^\al$ up to an overall constant. This constant must then be determined in
our case by
matching the series expansion of $\tet$ at infinity with (\ref{ez=}).
An important additional condition is that the eigenvalues of
$A_1,...,A_N, A_\infty$ (which are equal to the eigenvalues of
$L_1,...,L_N,L_\infty$) do not
differ by integers. In that case logarithmic singularities appear which we will
not include in
this paper.
It was Miwa, Sato and Jimbo \cite{Sato} who wrote down a solution to this
monodromy problem in
terms of correlators of free fermions with twistoperators at the positions of
the singularities.
It was Moore \cite{Moore} who translated it in the language of a conformal
field theory for free
fermions. Their solution was:
\be
Y^\al_\bt(z,z_o,\{a_i\})=(z_o-z)\frac{\langle V_\infty(\infty)\psi^\al(z)
\bar{\psi}_\bt(z_o)V_N(a_N)...V_1(a_1)\rangle}
{\langle V_\infty(\infty)V_N(a_N)...V_1(a_1)\rangle}
\label{correlator}
\ee
\begin{figure}[t]
\centerline{\psfig{figure=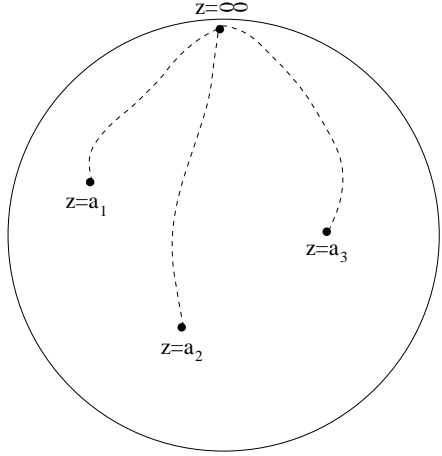,angle=-90,height=6cm}}
\caption{N Particles moving on the Riemann-sphere}
\label{Riemann-sphere}
\end{figure}
The operators $V_i$ are twistoperators to be defined shortly (\ref{twistop}).
First of all, it is easily seen that taking the limit $z\ra z_o$ we can write
\be
Y^\al_\bt(z,z_o,\{a_i\})=(z_o-z)\frac{\langle V_\infty(\infty)
\overbrace{\psi^\al(z)\bar{\psi}_\bt(z_o)}V_N(a_N)...V_1(a_1)\rangle}
{\langle V_\infty(\infty)V_N(a_N)...V_1(a_1)\rangle}~\label{exp-correlator}
\ee
+ holomorphic terms at $z=z_o$\\
Where we have denoted the Wick-contraction as:
\footnote{Note that this relation allows in principle for a more general set of
fields
$b^\al(z)$ and $c_\bt(z_o)$ which have spin $\lambda$ and $1-\lambda$
respectively (a b-c system).
We want however that the spin (conformal weight) of $\zeta^\al(z)$ is the same
as that of $b^\al(z)$,
so that they transform the same way under local transformations of the
coordinate $z$. This gives that
$\lambda=\frac{1}{2}$ as we have chosen in the text.}
\be
\overbrace{\psi^\al(z)\bar{\psi}_\bt(z_o)}=-\frac{\de^\al_\bt}{z-z_o}
\label{Wick}
\ee
Using the above formula we find that:
\be
Y^\al_\bt(z_o)=\de^\al_\bt
\label{Y(zo)}
\ee
as required by condition i). An important difference with \cite{Moore} is that
we have put a
twistoperator at $z=\infty$, altering the "outgoing vacuum" and changing the
scaling dimensions
of the fields at finite distances $a_i$.
The twist at infinity must reproduce the monodromy at infinity and contains the
information about
the total mass of the universe.\\
We now come to the subject of choosing the twistoperators. They must be
constructed in such a
way that they reproduce the correct monodromy properties at $z=a_i$. In the
following we will
consider a particle at $z=0$ with boost-matrix $B_i=B$ and twistoperator
$V_i(0)=V(0)$.
We will drop the index i in the following. The same story holds for all the
particles.
For the particle at $z=0$ we want:
\be
\psi^\al(e^{2\pi i}z)=M^\al_{~\bt}\psi^\bt(z)=(BRB^{-1})^\al_{~\bt}\psi^\al(z)
\label{mon-psi}
\ee
This means that the short distance behaviour of $\psi^\al(z)$ and $V(0) $ must
be of the form:
\be
V(0) \psi^\al(z)=\sum_{\bt }(B_i)^\al_{~\bt}z^{D_{\bt }}H^\bt(z)
\label{OPE-V-psi}
\ee
where B is a boost-matrix, $D_\bt $ are the eigenvalues of the matrix $L$ and
$H^\bt $ is a
holomorphic vector at $z=0$.
We can write this as:
\be
\sum_\bt  B^\al_\bt  z^{D_\bt}  H^\bt =\sum_\bt (B\Omega(z))^\al_\bt  H^\bt
\label{OPE-V-psi2}
\ee
where we supressed the index i and where:
\be
\Omega(z)=\left(\begin{array}{cc}
           z^{-\frac{\al}{2}}&0\\
           0&z^{\frac{\al}{2}}
            \end{array}\right)~~~~~~~~~\al=4Gm
\label{Omega}
\ee
Due to $\Omega(z)$ this expression is multi-valued and, by analytic
continuation,
we find if we move around the particle at $z=a_i$:
\be
(B\Omega(z))^\al_{~\bt }H^\bt \ra(BRB^{-1})^\al_{~\bt }(B\Omega(z)H)^\bt
\ee
where:
\be
R=\left(\begin{array}{cc}
       e^{-i\pi\al}&0\\
       0&e^{i\pi\al}
\end{array}\right)
\ee
This is of course precisely the right monodromy and our task is to find the
twistoprators
with this property. In ref.\cite{Moore} one finds:
\be
V =:\exp[-p_a\int_{C}dw~\bar{\psi}_\al t_{\al\bt}^a\psi_\bt]:
\label{twistop}
\ee
where $:\bar{\psi}_\al(w)t^a_{\al\bt}\psi_\bt(w):\equiv J^a$ is the
su(1,1) Kac-Moody current, and $C$ is a contour along the cut in the plane from
the position
$z=0$ to $\infty$, and $-p_at^a_{\al\bt}=L_{\al\bt}$. In order to obtain the
operator product
expansion (OPE) at $z=0$ we note that we can use the freedom to transform the
Dirac-spinors
$\psi^\al$ by a global SU(1,1) transformation.
\be
\psi^\al(z)\ra\tilde{\psi}^\al=\Lambda^\al_{~\bt}\psi^\bt(z)
\ee
Next we write:
\ba
V(0) \psi^\al(z)&=&:\exp[\int_{C}dw~\bar{\psi}^\mu
L_{\mu\nu}\psi^\nu]:\psi^\al(z)\\
&=&:\exp[\int_{C}dw~(\bar{\psi}B)^\mu(B^{-1}LB)_{\mu\nu}(B^{-1}\psi)^\nu]:\psi^\al(z)
\ea
Defining:
\ba
\tilde{\psi}^\al&=&(B^{-1}\psi)^\al\\
\tilde{\bar{\psi}}^\al&=&(\bar{\psi}B)^\al
\ea
we find:
\be
V(0) \psi^\al(z)=:\exp[\int_{C}dw~\tilde{\bar{\psi}}^\mu D_{\mu
\nu}\tilde{\psi}^\nu]:
B^\al_{~\bt}\tilde{\psi}^\bt(z)
\ee
Here $D_{\mu \nu}=p_0t^0_{\mu\nu}=\frac{\al}{2}t^0_{\mu\nu}$, $\al=4Gm$ and
$t^0_{\mu\nu}$=diag(1,-1).The next step is to bosonize the $\tilde{\psi}^\al$,
$\tilde{\bar{\psi}}^\al$
and $\tilde{V(0)}~(=V(0) )$. Of course we can not simultaneously diagonalize
all the $L_i$,
but we don't need that actually because we only need the OPE locally at $z=0$.
This result is
independent of the representation that we choose (i.e. fermionic or bosonic).
So we write:
\ba
\tilde{\psi}^\al(z)&=&:e^{i\varphi_\al(z)}:\\
\tilde{\bar{\psi}}(z)&=&:e^{-i\varphi_\al(z)}:\\
\sum_\al:\tilde{\bar{\psi}}^\al D_\al\tilde{\psi}^\al:&=& \sum_\al D_\al
\dd\varphi_\al\equiv \vec{D}\cdot\dd\vec{\varphi}\label{current}\\
\tilde{V}(0)&=&:e^{ i\vec{D}\cdot\vec{\varphi}(0)}:
\ea
Where $D_\al$ are the eigenvalues of $L$. Formally we should add Klein-factors
($c_\al$ and $\bar{c}_\al$) to the expression $:e^{i\varphi_\al}:$ and
$:e^{-i\varphi_\al}:$
to ensure that the fermions anticommute for all $\al$. We will leave them out
in the following
because we won't need them futher.
Note that if we bosonize the twistoperators (\ref{twistop}) according to the
rule (\ref{current}),
we also pick up a term at infinity:
$V(0)=e^{i\vec{D}\cdot\vec{\varphi}(0)}e^{-i\vec{D}\cdot\vec{\varphi}(\infty)}$.
This extra term is of course part of the twistoperator at infinity. There is
thus a
slight inconsistency in the notation. Because the twistoperators are
really nonlocal operators they can not be assigned to 1 point.\\
Using the bosonic representation and the following identity:
\be
:e^{i\vec{\kappa}\cdot\vec{\varphi}(z)}::e^{i\vec{\lambda}\cdot\vec{\varphi}(w)}:
=(z-w)^{\vec{\kappa}\cdot\vec{\lambda}}:e^{i\vec{\kappa}\cdot\vec{\varphi}(z)
+i\vec{\lambda}\cdot\vec{\varphi}(w)}:
\label{identity}
\ee
we find for the OPE:
\be
V(0) \psi^\al(z)=\sum_\bt B^\al_{~\bt}z^{D_\bt}H^\bt
\ee
where $H^\bt$ is again a holomorphic vector at $z=0$. This is precisely
expression
(\ref{OPE-V-psi}). So we esthablished the right behaviour of the solutions near
the points $a_i$.
The next question is obviously to check the behaviour of this solution at
infinity. To find that
one needs to know how the field $\psi^\al(z)$ scales, i.e. we need to know the
scaling dimensions
$\Delta^\al=2h^\al$ of the fermion. However the fact that we have put a twist
operator at infinity
spoils the naive scaling dimensions (see \cite{Fateev}).
Before we do that we like to get rid of the $z_o$ dependence. More precisely we
want to replace
condition i) in the beginning of this section by a different condition at
$z=\infty$. That new condition is:
\begin{itemize}
\item[i')] \be    \hat{Y}^0_\infty=I   \ee
\end{itemize}
To implement this new condition we first have to bring the field
$\bar{\psi}_\bt(z_o)$ to infinity,
using the rule:
\be
\bar{\psi}_\bt(\infty)=\lim_{z_o\ra\infty}\bar{\psi}_\bt(z_o)z_o^{\Delta_\bt}~~~~~~~\Delta_\bt=1
\label{brengoneindig}
\ee
In front of expression (\ref{correlator}) we have a factor $(z_o-z)$ and we use
this factor
($(z_o-z)\sim z_o$ if $z_o\ra\infty$) to take the limit (\ref{brengoneindig}).
However there
is already a twistoperator at infinity which implies that we will have to deal
with another
divergent factor. For convenience we will assume that the center of mass is not
moving, which
means that at infinity the monodromy is diagonal. The twist at infinity is
defined by the total
action of the twists at finite points. Because all these twists contain
integrals from the particle
position $a_i$ to infinity, and at infinity all these twists do not add up to
zero, there is a
nontrivial twist at infinity, see Fig. 4.\\
To see what happens if we bring the operator $\bar{\psi}_\bt(z_o)$ close to
this twistoperator
we first bring $V_\infty$ to a point $b$ in the finite plane and bosonize:
\be
V_\infty(b)\pbar=:e^{
-i\vec{D}_{\infty}\cdot\vec{\varphi}(b)}::e^{-i\varphi_\bt(z_o)}:\sim
(b-z_o)^{D_\bt^\infty}H_\bt(z)
\ee
where $H_\bt(z)$ is holomorphic at $z=b$. This suggests that if we want to take
the limit $z_o\ra b$
we have to do the following:
\be
\lim_{z_o\ra
b}(b-z_o)^{-D^\infty_\bt}V_\infty(b)\pbar=:V_\infty(b)\bar{\psi}_\bt(b):
\ee
One can check that if we take the limit:
\be
\lim_{z_o\ra b}(b-z_o)^{-D_\bt^\infty}Y^\al_{~\bt}
\ee
we have precisely the condition i') at z=b:
\be
\hat{Y}^0_b=I
\ee
This means that in taking the fermion $\pbar$ to infinity we need to multiply
by a factor $z_o^{D_\bt}$
in order to obey the new condition i') at infinity.
All together we write now:
\be
Y'^\al_{~\bt}=\lim_{z_o\ra\infty}z_o^{D^{\infty}_\bt}Y_{~\bt}^\al=
\frac{\langle:V_\infty\bar{\psi}_\bt:\p V_N...V_1\rangle}{\langle V_\infty
V_N...V_1\rangle}
\label{correlator'}
\ee
To determine the asymptotic behaviour of $Y'$ we have to calculate the scaling
behaviour of the $\p$.
To do that we again bosonize and note once more that we have chosen the center
of mass to be at rest,
meaning that $V_\infty$ has a "diagonal form". As mentioned before, the
operator $:V_\infty\bar{\psi}_\bt:$
at infinity spoils the naive scaling behaviour. In the bosonized picture we
have the following operator at infinity:
\be
O^\infty_\bt=:e^{-i\vec{D}_\infty\cdot\vec{\varphi}-i\varphi_\bt}:
\ee
Following (\cite{Fateev}) we must change the scaling dimensions by the
following rule:
\be
\Delta'^\al=1-(D^\infty_\al+\de^\al_\bt)
\ee
This means that the asymptotic behaviour of $Y'$ looks like:
\be
\left(\begin{array}{cc}
       z^{-\frac{\al_{tot}}{2}}H_{11}(z)&z^{-\frac{\al_{tot}}{2}-1}H_{12}(z)\\
       z^{\frac{\al_{tot}}{2}-1}H_{21}(z)&z^{\frac{\al_{tot}}{2}}H_{22}(z)
       \end{array}\right)
\ee
where $H_{ij}(z)$ are holomorphic functions as $z\ra\infty$. Next we choose
$\stackrel{o}{\zeta}^\al=(1,0)$ so that we have:
\be
\zeta^\al\stackrel{z\ra\infty}{\sim}\left(\begin{array}{c}
                              z^{-\frac{\al_{tot}}{2}}\\
                              z^{\frac{\al_{tot}}{2}-1}
    \end{array}\right)[b_\infty^\al+c_\infty^\al\frac{1}{z}+...]
 \ee
This is in fact the right asymptotic behaviour at $z\ra\infty$.
One can actually find an explicit expression for these correlators in
(\cite{Sato}).
This is a rather formal infinite series and hardly of any practical use to us.
Therefore we will not repeat it here. The first order matrix differential
equation can
be viewed as an equation for parallel
transport on a Riemann-sphere. We can define a connection:
\be
{\cal A}=\sum_i\frac{A_i}{z-a_i}
\label{ijkveld}
\ee
and a covariant derivative
\be
{\cal D}=\frac{\dd}{\dd z}-{\cal A}
\ee
In this picture $\tet$ is determined by parallel transporting the $e^\al(z_o)$
from $z_o$ to $z$. The result depends on the path we choose. Because the
connection ${\cal A}$ is holomorphic everywhere except at the singularities
$a_i$
and $\infty$ the result will only depend on the first homotopy class of the
punctured
sphere. In general the dependence of the $A_i$ on the points $a_i$ can be very
difficult.
If we use normalisation condition i) then also a dependence on $z_o$ is
possible.
Schlesinger determined a differential equation for this dependence by demanding
that if we move the particles to another location (for instance by
time-evolution!)
the monodromy will not change. This is because we don't allow the particles to
cross
a string connected to one of the other particles, by smoothly deforming the
string to
another position as a particle comes close to a string. This also means that
the topology
of the strings will become very important in the dynamics of the particles.
Another way of
putting the fact that the monodromy does not change during time-evolution is
that the flat
momenta $p_a$ are conserved. They act as a kind of charges in this picture. The
equations
found by Schlesinger are called isomonodromic deformation equations and are
given with the
normalisation i') by:
\footnote{If we choose normalisation i), the equations will also include a
$z_o$ dependence.
see \cite{Sato} and \cite{Moore}}
\ba
\frac{\dd Y'}{\dd a_i} &=& -\frac{A_i}{z-a_i}Y' \label{Sl1}\\
\frac{\dd A_i}{\dd a_j} &=& \frac{[A_j,A_i]}{a_j-a_i}~~~~~~~~~~~~~i\neq
j\label{Sl2}\\
\frac{\dd A_i}{\dd a_i} &=& -\sum_{k\neq i}
\frac{[A_k,A_i]}{a_k-a_i}\label{Sl3}
\ea
In order to get a feeling for the solution presented in (\ref{correlator}) we
will calculate
a simple example, namely the case where all $A_i$ commute. From
(\ref{Sl1},\ref{Sl2},\ref{Sl3})
we see that the $A_i$ do not depend on the particle's positions $a_i$. Next we
diagonalise all
$A_i$ simultaneously. This means that we are describing a static configuration
of particles.
It is very easy to find the exact solution to the differential equation
(\ref{dv}):
\be
Y=\prod_{i=1}^{N}(\frac{z-a_i}{z_o-a_i})^{D_i}
\ee
where :
\be
D_i=\left( \begin{array}{cc}
          \frac{-\al_i}{2}&0\\
          0&\frac{\al_i}{2}   \end{array}\right)~~~~~~~\al_i=4Gm_i
\ee
The answer in the normalisation i') is according to (\ref{correlator'}):
\be
Y'=\lim_{z_o\ra\infty}z_o^{D_\infty}Y=\prod_{i=1}^{N}(z-a_i)^{D_i}
\ee
In the static case it is indeed true that $M_{tot}=\sum_{i=1}^N m_i$.
Next we choose the vector $\stackrel{o}{\zeta}^\al=(1,0)$ so that we find:
\be
\zeta^\al=(\prod_{i=1}^{N}(z-a_i)^{-\frac{\al_i}{2}},0)
\ee
and
\be
\tet=(0,\prod_{i=1}^{N}(z-a_i)^{-\al_i},0)
\ee
Now using the formula (\ref{e=e*}) we can calculate the conformal factor
defined by $e^\phi=e^a_z(z)e_{\bar{z},a}(\bar{z})$ (see
(\ref{conformalmetric})):
\be
e^\phi=\prod_{i=1}^{N}(z-a_i)(\bar{z}-\bar{a_i})^{-\al_i}
\ee
which is the correct answer (see \cite{beginarticle})
On the other hand we would like to calculate the correlator
(\ref{correlator'}).
Because all $L_i$ commute we can bosonize the fermions at all points $a_i$
simulteneously.
We may view the fermion $\p$ as two twistoperators with "charge" $1$ and
$\pbar$ as 2
twistoperators with "charge" $-1$. The correlators that we have to calculate
($\al,\bt=1,2$)
are only nonzero if the total charge adds up to $\al_{tot}$. This is due to the
infrared cut-off
that has to be taken care off \cite{Fateev}. Only when the total charge is
$\al_{tot}$ we find a
nonzero correlator. Now using repeatedly formula (\ref{identity}) and noticing
that only for $\al=\bt$
the total charge adds up to $\al_{tot}$ we find:
\be
Y'=\frac{\prod_{i=1}^N(z-a_i)^{D_i}\prod_{i<j}^N(a_i-a_j)^{\vec{D_i}\cdot\vec{D_j}}}
{\prod_{i<j}^N(a_i-a_j)^{\vec{D_i}\cdot\vec{D_j}}}=\prod_{i=1}^N(z-a_i)^{D_i}
\ee
The use of the operators at infinity is now to ensure that the total charge
adds up to $\al_{tot}$:
\be
\sum_{i=1}^{N}\al_i=\al_{tot}~~~~~~~~~~~~~\al=\bt
\ee
We could call this a Coulomb-gas picture since it bears much resemblance with
the background
Coulomb-gas of ref.\cite{Fateev}. Finally we want to say something about the
correlator:
$\langle V_\infty V_N...V_1\rangle$. We can derive a kind of
Knizhnik-Zamalodchikov equation
for this correlator (see also \cite{Moore}). \footnote{In the mathematical
literature this is
called the tau-function.} For that consider first:
\ba
&&\frac{\frac{1}{2}\langle V_\infty[\pbar\frac{\dd}{\dd z}\p -\frac{\dd}{\dd
z_o}
\pbar\p] V_N...V_1\rangle}{\langle V_\infty V_N...V_1\rangle}\\
&&=-\frac{\de^\al_\bt}{(z_o-z)^2}+\frac{\langle V_\infty\tilde{T}^\al_\bt
V_N...V_1\rangle}
{\langle V_\infty V_N...V_1\rangle}
\ea
where we have defined:
\be
\tilde{T}_{\bt\al}\equiv :\frac{1}{2}\pbar\frac{\dd}{\dd
z_o}\psi_\al(z_o)-\frac{1}{2}
\frac{\dd}{\dd z_o}\pbar\psi_\al(z_o):
\ee
Because we have a background charge at infinity, the naive "energy-momentum
tensor"
$\tilde{T}\equiv tr\tilde{T}$ has to be corrected by a boundary term:
\be
T=\tilde{T}-\sum_{\al}\frac{1}{2}D^{\infty}_\al\dd:\bar{\psi}_\al\psi_\al:
\ee
The boundary term however does not change the simple pole term in the OPE so
that
we still have:
\be
L_{-1}=\oint_{a_i}dz\frac{T(z)}{2\pi i}=\oint_{a_i}dz\frac{\tilde{T}(z)}{2\pi
i}
=\frac{\dd}{\dd a_i}
\ee
So we find:
\be
\oint_{a_i}\frac{\langle V_\infty \tilde{T}V_N...V_1\rangle}{2\pi i\langle
V_\infty V_N...V_1\rangle}=\frac{1}{\langle V_\infty
V_N...V_1\rangle}\frac{\dd}
{\dd a_i}{\langle V_\infty V_N...V_1\rangle}
\label{Knizhnik-Zamalodchikov1}
\ee
On the other hand we may also write the correlator as:
\be
\frac{\frac{1}{2}\langle V_\infty [\pbar\frac{\dd}{\dd z}\p-\frac{\dd}
{\dd\bar{z}}\pbar\p] V_N...V_1\rangle}{\langle V_\infty V_N...V_1\rangle}=
-\frac{1}{2}\frac{\dd}{\dd z}\frac{Y_{\al\bt}}{z_o-z}+\frac{1}{2}\frac{\dd}
{\dd z_o}\frac{Y_{\al\bt}}{z_o-z}
\ee
This can be worked out by using an expansion around $z_o$:
\be
Y=I+{\cal A}(z_o)(z-z_o)+\frac{1}{2}[\frac{\dd}{\dd z_o}{\cal A}(z_o) +
{\cal A}^2(z_o)](z-z_o)^2
\ee
${\cal A}$ is defined in (\ref{ijkveld}). Integrating this expression around
$z_o=a_i$
and comparing with (\ref{Knizhnik-Zamalodchikov1}) we get:
\be
\frac{\dd}{\dd a_i}{\langle V_\infty
V_N...V_1\rangle}=\frac{1}{2}\oint_{a_i}dz_o~~
\sum_{i,j}\frac{tr~A_iA_j}{(z_o-a_i)(z_o-a_j)}
\label{Knizhnik-Zamalodchikov2}
\ee
As mentioned before there also exists a Schlesinger equation in which $z_o$ is
included
in the set deformation parameters (see \cite{Sato}). The equation we need is:
\be
\frac{\dd A_i}{\dd z_o}=\sum_j \frac{[A_i,A_j]}{z_o-a_j}
\ee
We can use this formula to find that $tr~A_iA_j$ does not depend on $z_o$ at
all!
This allows us to calculate the integral in (\ref{Knizhnik-Zamalodchikov2}) and
we
finally find:
\be
\frac{\dd}{\dd a_i} {\langle V_\infty V_N...V_1\rangle}=\frac{1}{2}\sum_{i\neq
j}
\frac{tr~A_iA_j}{a_i-a_j}{\langle V_\infty V_N...V_1\rangle}
\ee
Although this looks a bit like a Knizhnik-Zamalodchikov equation, there is are
differences.
Most importantly, the term $tr~A_iA_j$ may depend in a very complicated way on
the positions
of the particles $a_i$. The fact that we didn't find a
Knizhnik-Zamalodchikov-equation (with
constant matrices $A_i$) is interpreted in \cite{Ferrari} as a result of the
fact that
on the Cauchy-surface the translational symmetry is broken.\\
As solving the Riemann-Hilbert problem is equivalent to finding the $A_i$, we
have in general
not an explicit expression for them. Only in some special cases, where we are
able to find the
$A_i$ we can write down a full solution to the Riemann-Hilbert problem. It is
also very
interesting to look at the OPE of 2 twistoperators and determine the monodromy
if one
particle moves around the other. Alternatively one can look at the
exchange-algebra (or half-monodromy) and determine what happens if 2 particles
are
exchanged. We expect that these half-monodromies obey some non-abelian
representation
of the Yang-Baxter equation (see \cite{Verlinde}). We hope to say more about
this in the future.

\section{Discussion}
In a series of papers \cite{beginarticle,tHooft} 't Hooft solved the classical
N-particle
problem in 2+1 dimensional gravity. In his approach he used flat coordinates
$u^a$ that
obey unusual boundary conditions. As a result these coordinates become
multivalued.
In this polygon approach he developed a Hamiltonian formulation for the phase
space
variables\footnote{The phase space variables are the lengths of the edges of
the polygons and
the rapidities (boost-parameters) defined across the edges.} and by replacing
Poisson-Brackets
by commutators he was able to formulate a first quantized theory. It is however
rather difficult
to write down an explicit wave function in this formalism as all kinds of
transitions that could
take place have to be taken into account as boundary conditions on the wave
functions. Surely if
one wants to formulate a second quantized theory these boundary conditions
become untractable.
This was the main motivation for us to define a single-valued coordinate
system, much like in
the case of anyons where one has multi-valued and single-valued gauges. In the
case of anyons
the single-valuedness has a price, namely the introduction of an
electromagnetic  interaction
between the anyons. Also in the case of 2+1 dimensional gravity we can trade
the multivaluedness
for a nontrivial gravitational interaction.
This implies that we have to introduce a nontrivial metric $g_{\mu\nu}\neq
\eta_{\mu\nu}$.

 The physical quantities of the theory are the particle positions and the
conjugate momenta.
As the gravitational field doesn't have any degrees of freedom it is in
principle possible to
integrate them out. To do that one must first study the gravitational field in
our particular
gauge. That is precisely what we did in this article and we found that the 2
particle case could
be solved explicitly while in the N-particle case, the gravitational field (or
rather the dreibeins)
could be represented in terms of correlators of a two dimensional conformal
field theory.

In order to reduce the system to its physical degrees of freedom
we need to find a Hamiltonian. In section 1 it was argued that the Hamiltonian
is given by a
boundary term:
\be
H=\int d^2x~~\sqrt{h}~^{(2)}R=\oint dz~\dd\phi(z\bz)~~+~~\oint
d\bz~\bar{\dd}\phi(z,\bz).
\ee
where $\phi$ is defined by: $e^\phi=e^a_z(z)e_{\bar{z},a}(\bar{z})$. This
expression can be
simply evaluated by using the monodromy-matrix at infinity and the asymptotic
behaviour of
$\phi(z,\bz)$ (\ref{line-element3}). One finds the expected result;
$H=\al_{tot}$, the total
deficit angle of the universe. By using (\ref{Mtotaal}) it
then easy to find the well known result:
\be
-2\pi iH~t^o_{\al\bt}=\log[\prod_{i=1}^{N}e^{2\pi iL_i}]_{\al\bt}
\ee
where $L_i=-p^{i}_a t^a_{\al\bt}$ ; $p_a^{(i)}$ are the flat momenta and
$t^a_{\al\bt}$ are the
generators of SO(2,1) (or equivalently SU(1,1)). Because we restricted
ourselves in this paper
to particles with positive mass, and didn't consider anti-particles with
negative mass (which
could in principle be defined) we have positive total energy. It is however
also particular
simple in our gauge to prove that for arbitrary matter couplings (that obey
reasonable boundary
conditions at infinity) the total energy of an open universe is positive
definit (see \cite{Ash,Henneaux}).
This is of course a nessecary property in order to have stable vacuum in a
second quantized theory.
It is also important to realise that gravity in 2+1 dimensions contains long
range interactions.
This can be seen most easily in the flat coordinate-system
because the wedge cut out of space-time extends out to infinity. This means
that we don't have an
asymptotic flat region. If we want to define an S-matrix in the future we know
that we will
have to define in- and out-states with great care.
We hope to say more about the particle dynamics and the quantisation of this
system in
the future.

\section*{Acknowledgements}
I would like to thank G. 't Hooft, E. Verlinde, H. Verlinde and K. Sfetsos
for interesting discussions during the preparation of this work.

\section*{Note added}
While writing this paper I found out that A. Bellini et al. \cite{Bellini}
published
a preprint concerning the same issues we adress.

\end{document}